\documentclass[a4paper, 10pt, twocolumn]{article}

\usepackage{amsfonts,amssymb,amsmath,amsthm}
\usepackage{subfigure}
\usepackage{graphicx}
\usepackage[footnotesize]{caption}

\input{alphabet.tex}

\usepackage[top=1.5cm, left=1.5cm, right=1.5cm, bottom=1.5cm]{geometry}

\title{High-resolution three-dimensional crystalline microscopy}

\author{Marc ALLAIN$^{1}$, Virginie CHAMARD$^1$ and Stephan O.  HRUSZKEWYCZ$^{2}$.\\
\footnotesize $^1$Aix-Marseille Univ, CNRS, Centrale Marseille, Institute Fresnel, 13013 Marseille, France.\\ 
\footnotesize $^2$ Materials Science Division, Argonne National Laboratory, Lemont, IL 60439, USA.
}
\date{\empty} 

\renewenvironment{abstract}{\bf\small {\em\ Abstract---}}{}

\begin{document}

\maketitle

\begin{abstract}
In this communication, we discuss how 3D information about 
the structure of a crystalline sample is encoded in  Bragg 
3DXCDI measurements. Our analysis brings to light the role of 
the experimental parameters in the quality of the final reconstruction. 
One of our salient conclusions is that these parameters can be set 
prior to the ptychographic 3DXCDI experiment and that the spatial 
resolution limit of the 3D reconstruction can be evaluated accordingly.
\end{abstract}

\section{Introduction}
\label{sec:introduction}

Since its introduction in the early 2000’s \cite{ref1,ref2}, three-dimensional X-ray 
Coherent Diffraction Imaging (3DXCDI) has widely demonstrated its ability 
to provide non-destructive three-dimensional (3D) images of complex 
nanostructures. Two key features of 3DXCDI are noteworthy: 1) 3DXCDI 
offers the possibility to measure data either in a Bragg \cite{ref2} or in a forward 
geometry, the former case providing 3D images of strains in crystalline 
materials \cite{ref3,ref4}; and 2) 3DXCDI can be executed with a ptychographic 
(spatial) scan, hence providing images of extended samples \cite{ref5,ref6,ref7}. For 
these reasons, 3DXCDI opens a wide playground for x-ray microscopy.
%
%
In this communication, we discuss how 3D information about 
the structure of a sample is encoded in Bragg 3DXCDI measurements. 
In particular, our analysis brings to light the role of the experimental 
parameters in the quality of the final reconstruction. 

%

\section{Spatial scan at a Bragg peak}
\label{sec:first-section}


Any 3DXCDI approach is inherently a \textit{lens-less tomographic} modality: it is “lens-less” 
because the dataset (a series of coherent intensity patterns) is numerically 
inverted by a phasing algorithm, and “tomographic” since each collected 
diffraction intensity is drawn from the sample \textit{via} a tomographic 
measurement. 

\begin{figure}[t]
  \centering
	  \includegraphics[height=0.5\textwidth]{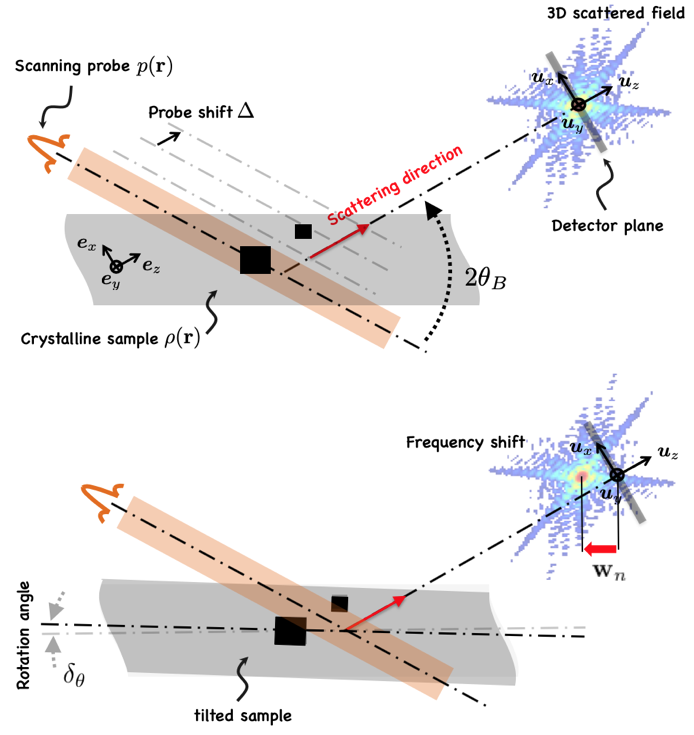}
  \caption{%
    {\bf A simplified Bragg ptychography experiment.} (Upper) When the incoming beam 
and the scattering direction define a specific Bragg angle $\theta_B$,  the Bragg condition 
for the chosen ($H$,$K$,$L$) Bragg peak is met and a diffraction (intensity) signal shows 
up at the detection plane. As the sample is shifted in a focused probe, the diffraction signal 
is recorded in each spatial position.  (Lower) A series of spatial scan can be performed 
with various tilts of the sample. This result in more 3D spatial information extracted from the sample.}
  \label{fig1}
\end{figure}

Some details about the Bragg ptychographical experiment  are now provided.
Let us introduce first the \textit{exit-field} 
\begin{equation}
\label{farfield}
\psi_m = p_m \times \rho
\end{equation}
where $m\in \{0,\cdots M-1\}$ is the position index during the spatial 
(ptychographical) scan. In the relation above, $\rhob:\eR^3 \rightarrow \eC$ denotes 
the \textit{scattering density} \cite[Sec. 7.1.2]{ref11} of the diffracting crystal and  
$p_m:\eR^3 \rightarrow \eC$ is the $m$-th coherent \textit{probe} 
illuminating the sample. When a Bragg condition is met\footnote{%
According to the usual convention in coherent Bragg diffraction, 
the origin of reciprocal space $(q_x=0, q_y=0, q_z=0)$ corresponds to 
a reciprocal space Bragg peak denoted by the reciprocal space vector 
$\textbf{G}_{\textit{HKL}}$ for a given $(H,K,L)$ Bragg reflection, see for instance \cite{ref8}.
}, both quantities are conveniently expressed with coordinates 
$\rb:=(\rb_\bot,r_z)\in \eR^3$ within a 3D frame in (direct-)space 
matching the detection geometry, see Fig.~1. In addition, for the sake of 
simplicity, we consider that the probing field is generated from a probe 
function $p$  shifted along the \textit{scattering 
direction} $\eb_z$ only\footnote{%
In general, the 3D shifting 
of the probe can be casted within a non-orthogonal frame $(\eb_x', \eb_y, \eb_z)$  
with $\eb_x'$ pointing along the direction of the incoming probe, see Fig.~\ref{fig1}. In the 
x-ray regim,  the probe $p(\rb)$ is invariant along $\eb_x'$ and any shift
along this direction does not provide any spatial information about the sample.   
}
\textit{i.e.,} we have
%
$
p_m(\rb):= p(\rb - \rb_m)
$
%
with
 %
$
\rb_m :=  0\times\eb_x + 0\times \eb_y +  m\Delta \eb_z
$
%
where $\Delta\in \eR$ is the step size of the spatial scan. 
In Bragg ptychography, as in any coherent diffraction method, the 
measurement is the intensity of the scattered-field collected by 
an array detector. Under the first-order born approximation 
\cite[Sec. 8.10.1]{ref9}, the scattered wave-field collected in the ``far-field'' reads 
at the detector plane   
\begin{equation}
  \label{diffracted_field}
  \Psi(\qb_\bot; r_z = m\Delta) \, =\,  \widetilde{\psi}_m(\qb_\bot,q_z=0) 
\end{equation}
where $\tilde{\psi}_m$ is the 3D Fourier transform of $\psi_m$ and $\qb:=(\qb_\bot,q_z)$
is the 3D  frequency (or \textit{reciprocal-space})  coordinates. Finally, the 
expected (\textit{i.e.,} noise-free) measurements at the detector plane are 
given by the intensity of the scattered field  \eqref{diffracted_field}
\begin{equation}
   \label{intensity}
    I (\qb_\bot ; r_z = m\Delta)  \,=\, | \Psi (\qb_\bot; r_z = m\Delta) |^2.
 \end{equation}
%
Although the relations \eqref{diffracted_field} and \eqref{intensity} are useful 
in deriving actual Bragg ptychography 
reconstruction algorithms (see for instance \cite{ref6,ref7}), it does not tell anything 
about the spatial information extracted from the sample \textit{via} 
the ptychographical measurements.  
A substancial leverage to address this question is provided by the following result, 
easily deduced from  the Slice Projection Theorem \cite[Sec 6.3.3]{ref10} 
\begin{equation}
  \label{diffracted_field_FCT_}
  \Fc_\bot^{-1}\Psi \, =\,  \rho \otimes_{\text{//}} p. 
\end{equation}
In the relation above, $\Fc_\bot$ is the bi-dimensional (2D) Fourier 
transform with respect to $\rb_\bot$, and $\otimes_{\text{//}}$ 
is the one-dimensional convolution operator acting along the 
scattering direction $\eb_z$. In other words, the 3D quantity
\begin{equation}
  \label{diffracted_field_FCT}
  g(\rb) \, :=\,  (\rho \otimes_{\text{//}} p)(\rb) 
\end{equation}
is an approximation of the scattering-density 
$\rho$ built on a filtering by the probe profile along $\eb_z$. 
The scattering-density approximation $g$ given in \eqref{diffracted_field_FCT} 
appears in a previous publication from the authors \cite{ref12}. In this communication, 
this relation is a pivotal tool in deriving \textit{resolution limits} and \textit{sampling 
conditions} for Bragg ptychography experiments.  

Let us assumed that the probe profile along $\eb_z$ is a 
band-limited function with its support strictly contained 
in the domain $\Omega_z:= [-\bar{q}_z, \bar{q}_z]$ with $\bar{q}_z 
\geq 0$. 
We deduce  from \eqref{diffracted_field} and \eqref{diffracted_field_FCT} 
that the scanning step-size $\Delta$ should be set \textit{at least} such 
that  the spatial information is preserved in $g(\rb_\bot, r_z = m\Delta)$, 
$m\in \eN$. The Shannon-Nyquist sampling rate is then driven by the  
maximal frequency $\bar{q}_z$ in the (assumed) band-limited probe 
profile. In addition, because the spatial sampling is performed over 
\textit{wave-field intensities} \eqref{intensity},     it is not difficult to 
show that the sampling rate should be at least twice the Shannon-Nyquist 
limit for the wave-field, \textit{i.e.,}  the following condition 
\begin{equation}
  \label{samping_rate}
  \Delta > 1/(4 \bar{q}_z) 
\end{equation}
ensures that the approximation $g$ given by  \eqref{diffracted_field_FCT} 
can be retrieved from the series of (noise-free) intensity measurements 
\eqref{intensity}. The resolution bounds one may achieve in practice are 
also provided by  $g$, as this latter function is the best 3D approximation 
of the scattering density  one can expect from the spatial scan. Because the 
convolution in \eqref{diffracted_field_FCT} acts as a pointwise multiplication with respect to 
the coordinates $\rb_\bot$,  the spatial resolution bounds along the directions 
$\eb_x$ and $\eb_y$ are not restricted by the probe. In the third direction $\eb_z$,  
the spatial resolution is restricted by the convolution kernel in 
\eqref{diffracted_field_FCT}, leading to the bound 
\begin{equation}
  \label{resolution}
  R_z = 1/\bar{q}_z. 
\end{equation}
The so-called  ``rocking-curve'' can nevertheless extend further this resolution 
limit \textit{via} angular diversity; this topic is developed in the next section.     

\section{Additional angles}
\label{sec:second-section}

When Bragg 3DXCDI was introduced in the early 2000's\footnote{%
Bragg 3DXCDI was introduced as a natural extension of standard CDI techniques.
In this context, the method aimed at imaging isolated nano-crystals, with restricted 
supports small enough so that the unfocused coherent beam illuminating the sample
can be considered as a single plane wave. The method was then mostly understood as 
a 3D Fourier synthesis strategy \cite{ref2,ref3}. 
}, the method relied on 
the \textit{sample rotation} to explore the 3D Fourier components of the sample 
to retrieve. In this context, a single 3D Bragg peak is probed by the camera plane 
while the sample is tilted. It is nevertheless a very peculiar tomographic 
modality: as the Bragg peak sits at a given point in the 3D reciprocal lattice of 
the probed crystal, the whole 3D Bragg peak is probed with unusually small\footnote{%
If the chosen (probed) Bragg peak is \textit{not} the one that sits at the origin 
of 3D reciprocal lattice, the angular range required for a full 3D scan 
is $\sim 1^\circ$.    
}  angular ranges. In addition, the sample rotation results in a \textit{cartesian}, rather 
than \textit{polar} sampling of the Bragg peak, see Fig.~1-Lower. 
When Bragg 3DXCDI is performed with a scanning (focused) probe, the relation 
\eqref{diffracted_field_FCT} clearly states that a  3D reconstruction can still be 
obtained without sample rotation. If the sample rotation is \textit{also} performed, 
we can expect \textit{more} spatial information to be extracted. This is the question 
we aim at addressing in this section.  

In Bragg geometry, 
a small rotation of the sample, by an angle $\delta_\theta$, results in a frequency 
shift by $\textbf{w}$ of the 3D Fourier transform of the scattering-density 
$\rho$, see Fig.~1-Lower. The scattered field at the camera plane reads then 
\begin{equation}
  \label{diffracted_field_tilted}
  \Psi_{\textbf{w}}(\qb_\bot; r_z = m\Delta) \, =\,  \widetilde{\psi}_{m; \textbf{w}}(\qb_\bot,q_z= 0) 
\end{equation}
where $\psi _{m; \textbf{w}} (\rb) := p_m (\rb) \times \rho (\rb) \, e^{j 2\pi \textbf{w}^t\rb}$ is the 
\textit{modulated} exit-field. The relation \eqref{diffracted_field_FCT_} is modified 
accordingly 
\begin{equation}
  \label{diffracted_field_FCT_tilted_}
  \Fc_\bot^{-1}\Psi_{\textbf{w}} \, =\,  (\rho \times e^{j 2\pi \textbf{w}^t\bullet}) \otimes_{\text{//}} p.
\end{equation}
The relation above shows that the accessible frequency domain along $\ub_z$ is now 
$\Omega_z (\textbf{w})  := \Omega_z \oplus \textbf{w}^t \ub_z$ (where $\oplus$ is 
the \textit{Minkowski sum}).
In practice, a series of $N$ tilts is usually performed, inducing an equivalent series of 
frequency shifts denoted $\Wc := \{\textbf{w}_n\}_{n=1}^N$. The best approximation one 
can achieve is then of the form \eqref{diffracted_field_FCT} and reads 
\begin{equation}
  \label{diffracted_field_FCT_tilted}
  g(\rb\,;\,\Wc) \, =\,  [\rho \otimes_{\text{//}} p(\cdot \,;\,\Wc) ](\rb) 
\end{equation}
where the \textit{equivalent} probe
%
$
  p(\rb \,;\,\Wc) := \sum_n p (\rb) \times e^{j 2\pi \textbf{w}_n^t\rb}
$
%
defines the spatial information extracted from the joint spatial/angular scan. The set 
of frequency that are extracted by this equivalent probe are $\Omega_z (\Wc)  := 
\Omega_z \oplus \sum_n \textbf{w}_n^t \ub_z$, and the resolution limit 
along  $\eb_z$ is obviously better than \eqref{resolution} and reads  
\begin{equation}
  \label{resolution_tilted}
\textstyle
  R_z' = 1/\bar{q}_z' \quad \text{with} \quad \bar{q}_z' :=  \bar{q}_z  + 
  \sum_n |\textbf{w}_n^t \ub_z|.
\end{equation}
We underline that the sampling condition \eqref{samping_rate} is actually 
unchanged when angular diversity is considered. A remaining, potential 
issue is that $\Omega_z (\Wc)$ may not be a compact set, hence creating 
un-probed ``holes'' in the 
frequency space of the approximation \eqref{diffracted_field_FCT_tilted}.  
The condition $||\textbf{w}_n|| \cos \theta_B \leq \bar{q}_z, \forall n,$
with $\theta_B$ the Bragg angle nevertheless ensures a continuous probing 
of the frequency domain.
%
%

This last section clearly connects Bragg ptychography to other super-resolved 
imaging techniques, \textit{e.g.,} \textit{structured illumination microscopy} \cite{ref13}, 
\textit{synthetic aperture} \cite{ref14} strategies. We also stress that these resolution limits 
are reached only in the asymptotic limit, with noise-free intensity measurements.  
In practice, both the photon shot-noise and the physical extension of the camera will 
reduce the effective resolution in all three directions.   

\section*{Acknowledgement}
This work was supported by the European Research Council (European Union's 
Horizon H2020 research and innovation program grant agreement No 724881).
Work at Argonne National Laboratory (development of the Bragg ptychography 
forward model) was supported by the U.S. Department of Energy (DOE), Office 
of Basic Energy Sciences (BES), Materials Science and Engineering Division.


\end{document}